\documentclass[12pt]{article}

\usepackage{latexsym}
\usepackage{tikz}
\usetikzlibrary{arrows,automata}
\usepackage{graphicx}
\usepackage{dcolumn}

\usepackage{amssymb}
\usepackage{amsmath}

\newcommand{\bq}{\begin{equation}}
\newcommand{\eq}{\end{equation}}

\newcommand{\bqr}{\begin{eqnarray}}
\newcommand{\eqr}{\end{eqnarray}}

\newcommand{\bqrx}{\begin{eqnarray*}}
\newcommand{\eqrx}{\end{eqnarray*}}

\newcommand{\br}{\begin{array}}
\newcommand{\er}{\end{array}}

\newcommand{\blsk}{\baselineskip}

\begin{document}

\pagestyle{plain}
\pagenumbering{arabic}

\setlength{\parindent}{0in}
\setlength{\parskip}{3ex}
\setlength{\footskip}{.6in}



\setlength{\blsk}{.25in}%
\vspace*{.6in}
\begin{center}
Quantum walks in the density operator picture
\end{center}
\begin{center}
Chaobin\, Liu  \footnote{cliu@bowiestate.edu, Department of Mathematics, Bowie State University\\Bowie, MD 20715 USA}

\end{center}

\begin{abstract}
{A new approach to quantum walks is presented. Considering a quantum system undergoing some unitary discrete-time evolution in a directed graph G, we think of the vertices of G as sites that are occupied by the quantum system, whose internal state is described by density operators. To formulate the unitary evolution, we define reflections in the tensor product of an internal Hilbert space and a spatial Hilbert space. We then construct unitary channels that govern the evolution of the system in the graph. The discrete dynamics
of the system (called quantum walks) is obtained by iterating the unitary channel on the density operator of the quantum system. It turns out that in this framework, the action of the unitary channel on a density operator is described by the usual matrix multiplication.
}
\end{abstract}%

\section{Introduction}

Considering a quantum system undergoing some discrete-time evolution in a directed graph G, we think of the vertices of G as sites that are occupied by the quantum system, whose inner state is determined by density operators. The quantum system is assumed to be isolated and closed. According to the principle of quantum mechanics, the evolution of the quantum system is described by a unitary transformation \cite{NC2000}. To the best of our knowledge, none of the existing quantum schemes or quantum frameworks (such as quantum walks and open quantum walks) exactly models the evolution of the quantum system in this scenario. 
\vskip 0.1in
Specifically, even though open quantum walks \cite{APS12, APSS12, G2008} do concern density operators, their transformations are usually not unitary since they are formulated as unital, trace-preserving and completely positive maps on graphs. The frameworks of discrete-time quantum walks in \cite{ABNVW01, W2001, AAKV01, K03, Konno05, K06, VA12} and the references cited therein are used to describe the unitary evolution of a quantum system, but the quantum evolutions address pure states of the system. It is noted that the unitary maps in the frameworks are defined by conditional shifts and coin operators. There are two other notable schemes for unitary evolution of quantum walks \cite{S04} and \cite{PRR05}, where the unitary transformations also deal with pure states of the quantum system. In \cite{S04}, the unitary transformation is defined by a swap operator and a reflection operator, while in \cite{PRR05}, the unitary map is given by two consecutive reflection-like operators. It may be of interest to read the lecture note \cite{W12} regarding the relationship between these unitary transformations. 

\vskip 0.1in
In this note, we aim to introduce an approach to the evolution of the aforesaid quantum system in the density operator picture. Enlightened by the idea presented in \cite{S04}, we will employ reflection operators in an enlarged Hilbert space to construct unitary transformations which govern the evolution of the system. Following convention, we will also call the evolution of this system quantum walks. 


\section{General setup for quantum walks in the density operator picture}

We may assume that the unitary transformation of a quantum system takes place on a directed graph $G(V, E)$. Here $V$ is the set of the vertices of $G$, $E$ is the set of the oriented edge of $G$, $E=\{(j,k): j,k \in V\}$.  Let $\mathcal{H}_V$ be a complex Hilbert space with the standard orthonormal basis $\{|j\rangle\}_{j\in V}$. The set of bounded linear operators on $\mathcal{H}_V$ is denoted by $\mathfrak{B}(\mathcal{H}_V)$. Let $\mathcal{H}_C$ be an $n$-dimensional complex Hilbert space, spanned by the set of degrees of freedom $C=\{c_1, c_2, ..., c_n\}$, the set of bounded linear operators on $\mathcal{H}_C$ is denoted by $\mathfrak{B}(\mathcal{H}_C)$.

\vskip 0.1in

For a Hilbert space $\mathcal{H}$, let $\mathfrak{B}(\mathcal{H})$ be the set of bounded linear operators on $\mathcal{H}$ with the {\it Hilbert-Schmidt} inner product, defined by 
$$\langle A,B\rangle=\mathrm{tr}(A^{\dagger}B).$$
It can be verified that $\mathfrak{B}(\mathcal{H})$ becomes a Hilbert space with this inner product.

\vskip 0.2in

Associated with the quantum system on $G$ is the Hilbert space $\mathcal{H}_C\otimes \mathcal{H}_V^{\otimes 2}$, which is called the {\it state space} of the system. The system is completely described by its {\it state vector}, which is a unit vector in the system's state space. For instance, a typical state vector may be of the form $u\otimes |lm\rangle$ where $u\in \mathcal{H}_C$ as an $n$-dimensional column vector. In the density operator picture, the Hilbert space becomes $\mathfrak{B}(\mathcal{H}_C)\otimes \mathfrak{B}(\mathcal{H}_V\otimes\mathcal{H}_V)$, and the system is completely described by a positive operator with trace one (called a {\it density operator}), acting on the state space of the system. For example, a typical density operator may be written in the form of $\rho\otimes |lm\rangle\langle lm|$ where $\rho\in \mathfrak{B}(\mathcal{H}_C)$, which is a positive operator with $\mathrm{tr}(\rho)=1$.

\vskip 0.2in

The quantum system described in the aforesaid framework can be understood as a system consisting of a particle that has an internal quantum state ($u$ or $\rho$) and occupies a classical state ($|lm\rangle$ or $|lm\rangle\langle lm|$). This particle evolves from one state to another under certain unitary transformation to be defined as follows.
\vskip 0.2in

To define the unitary transformation, we first introduce the states in the Hilbert space $\mathcal{H}_C\otimes \mathcal{H}_V^{\otimes 2}$ :
\begin{equation}
|\psi_j\rangle:=\sum_k v_j^k\otimes |j,k\rangle,\label{key}
\end{equation}
where the column vectors $v_j^k$ satisfy the unital condition condition, i.e., $\sum_k(v_j^k)^{\ast}v_j^k=1$.  $v^{\ast}$ is the Hermitian transpose of $v$. 

\vskip 0.2in
It is noted that the pairs $(j,k)$ in Eq.(\ref{key}) indicate the edges of $G$ directed from the vertex $j$ since the unitary transformation on $G$ is expected to be subject to the geometric structure of $G$.

\vskip 0.2in

Secondly, we define 
\begin{equation}
\Pi:=\sum_j^{|V|}|\psi_j\rangle \langle \psi_j|, \label{}
\end{equation}
which is the projection on $\mathrm{span}\{|\psi_j\rangle:j=1, 2,..., |V|\}$, denoted by $\mathcal{H}_{\psi}$. This is a subspace of the augmented Hilbert space $\mathcal{H}_C\otimes\mathcal{H}_V^{\otimes 2}$. It can be verified that $2\Pi-1$ is a reflection operator on $\mathcal{H}_C\otimes\mathcal{H}_V^{\otimes 2}$. 

\vskip 0.2in
Then let us define 
\begin{equation}
S:=\mathbb{I}_c\otimes\sum_{j,k=1}^{|V|}|j,k\rangle\langle k,j| \label{}
\end{equation}
to be the operator that exchanges the two registers. Here $\mathbb{I}_c$ is the identity operator on $\mathfrak{B}(\mathcal{H}_C)$.
\vskip 0.2in

Finally, we define 
\begin{equation}
U:=S(2\Pi-\mathbb{I}).\label{}
\end{equation}
Here $\mathbb{I}$ is the identity operator on $\mathfrak{B}(\mathcal{H}_C)\otimes \mathfrak{B}(\mathcal{H}_V\otimes\mathcal{H}_V)$. It is noted that $U\in \mathfrak{B}(\mathcal{H}_C)\otimes \mathfrak{B}(\mathcal{H}_V\otimes\mathcal{H}_V)$, and $U$ is a unitary operator.

\vskip 0.2in

For $v\in \mathcal{H}_C\otimes\mathcal{H}_V^{\otimes 2}$, we think of $v$ as a column
vector and define $Uv$  by the usual matrix multiplication. It can be shown that
$Uv$ is again a vector state in $\mathcal{H}_C\otimes\mathcal{H}_V^{\otimes 2}$. For $\tau\in \mathfrak{B}(\mathcal{H}_C)\otimes \mathfrak{B}(\mathcal{H}_V\otimes\mathcal{H}_V)$, we define $U\tau$ and $\tau U^{\dagger}$ by the usual matrix multiplication. It is evident that $\mathrm{tr}(U\tau U^{\dagger})=\mathrm{tr}(\tau).$ 






\vskip 0.2in

With the operators defined above, a single step of the quantum walk is defined as $\Phi(\rho)=U\rho U^{\dagger}$. For a given $\rho_0\in \mathfrak{B}(\mathcal{H}_C)\otimes \mathfrak{B}(\mathcal{H}_V\otimes\mathcal{H}_V)$ satisfying  $\mathrm{tr}(\rho_0)=1$, the expression $\rho_t=\Phi^t(\rho_0)$ is called the state of the walk at time $t$. The corresponding quantum walk with the initial state $\rho_0$ is represented by the sequence $\{\rho_t\}_{t=0}^{\infty}$. Then the sequence of time iterated states models the temporal evolution of the walk in the density operator picture.

\vskip 0.2in
To formulate the probability distribution of the quantum walks, we introduce quantum measurements for the quantum walks. The quantum measurements are given by a sequence of {\it quantum effects}, denoted by $E_{jk}=\mathbb{I}_c\otimes |j,k\rangle \langle j,k|$ satisfying $\sum_{jk}E_{jk}=\mathbb{I}$.

\vskip 0.2in

If the state of the quantum system is $\rho$ immediately before the measurement, then the probability that the effect $E_{jk}$ occurs (has a yes answer) is given by $P(E_{jk}) = \mathrm{tr}(E_{jk}\rho)$. The state of the system after the measurement is  
$$\frac{E_{jk}\rho E_{jk}}{\mathrm{tr}(E_{jk}\rho E_{jk})}.$$
The probability of finding a walker at the position $|j\rangle$ is $P(j)=\sum_kP(E_{jk})$. 

\vskip 0.2 in

\begin{figure}
\begin{center}
\begin{tikzpicture}[->,>=stealth',shorten >=1pt,auto,node distance=1.8cm,
semithick]
\tikzstyle{every state}=[fill=red,draw=black,thick,text=white,scale=1]
\node                      (al)                {$\cdots$};
\node[state]         (A)  [right of=al]             {$-2$};
\node[state]         (B) [right of=A] {$-1$};
\node[state]         (C) [right of=B] {$0$};
\node[state]          (D) [right of=C] {$1$};
\node[state]          (E) [right of=D]  {$2$};
\node                   (ar)     [right of=E]           {$\cdots$};
\path (al)     edge                        node        {}              (A);
\path (A) edge   node[above] {} (B);
\path (B) edge   node[below] {} (C);
\tikzstyle{every state}=[fill=blue,draw=black,thick,text=white,scale=1]
\path (C) edge  node[left] {} (D);
\path (D) edge  node[right] {} (E);
\path (E) edge node {} (ar);
\path (A) edge node {} (al);
\path (B) edge  node {} (A);
\path (C) edge  node {} (B);
\path (D) edge  node {} (C);
\path (E) edge   node {} (D);
\path  (ar) edge node {} (E);
\end{tikzpicture}
\end{center}
\caption{Quantum walks on the line}  \label{fig:QWonLine}
\end{figure}
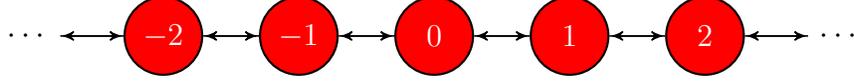

To illustrate our framework described above, we study quantum walks on one-dimensional infinite lattice (see Figure \ref{fig:QWonLine}) with the defining states given in Eq.(\ref{key}),
\begin{equation}
|\psi_j\rangle:=|[\frac{-i}{2},\frac{1}{2}]^T\rangle \otimes |j,j+1\rangle+|[\frac{1}{2},\frac{1}{2}]^T\rangle \otimes |j,j-1\rangle. \label{}
\end{equation}

Then we have \begin{eqnarray}
|\psi_j\rangle\langle \psi_j|=\left[\begin{array}{cc}
\frac{1}{4}& \frac{-i}{4}\\
\frac{i}{4}    & \frac{1}{4}
\end{array}\right]\otimes |j,j+1\rangle\langle j,j+1|+
\left[\begin{array}{cc}
\frac{-i}{4}& \frac{-i}{4}\\
\frac{1}{4}    & \frac{1}{4}
\end{array}\right]\otimes|j, j+1\rangle\langle j,j-1|   \nonumber \\ 
+\left[\begin{array}{cc}
\frac{i}{4}& \frac{1}{4}\\
\frac{i}{4}    & \frac{1}{4}
\end{array}\right]\otimes |j,j-1\rangle\langle j,j+1|+
\left[\begin{array}{cc}
\frac{1}{4}& \frac{1}{4}\\
\frac{1}{4}    & \frac{1}{4}
\end{array}\right]\otimes|j, j-1\rangle\langle j,j-1|.\, \label{}
\end{eqnarray}

For the initial state \begin{equation}
\rho_0= \left[\begin{array}{cc}
\frac{1}{2}& \frac{1}{2}\\
\frac{1}{2}    & \frac{1}{2}
\end{array}\right]\otimes|0,1\rangle\langle 0,1|,\label{}
\end{equation}

one step of evolution leads to the state 
\begin{eqnarray}
\rho_1=\Phi(\rho_0)=\left[\begin{array}{cc}
\frac{1}{4}& \frac{i}{4}\\
\frac{-i}{4}    & \frac{1}{4}
\end{array}\right]\otimes |1,0\rangle\langle 1,0|+
\left[\begin{array}{cc}
\frac{-1}{4}& \frac{-i}{4}\\
\frac{-1}{4}    & \frac{-i}{4}
\end{array}\right]\otimes|-1, 0\rangle\langle 1,0|   \nonumber \\ 
+\left[\begin{array}{cc}
\frac{-1}{4}& \frac{-1}{4}\\
\frac{i}{4}    & \frac{i}{4}
\end{array}\right]\otimes |1,0\rangle\langle -1,0|+
\left[\begin{array}{cc}
\frac{1}{4}& \frac{1}{4}\\
\frac{1}{4}    & \frac{1}{4}
\end{array}\right]\otimes|-1,0\rangle\langle -1,0|;\, \label{}
\end{eqnarray}

Two consecutive steps of evolution lead to the following state
\begin{eqnarray}
\rho_2=\Phi(\rho_1)=\left[\begin{array}{cc}
\frac{1}{8}& \frac{-i}{8}\\
\frac{i}{8}    & \frac{1}{8}
\end{array}\right]\otimes |2,1\rangle\langle 2,1|+
\left[\begin{array}{cc}
\frac{1}{8}& \frac{-1}{8}\\
\frac{i}{8}    & \frac{-i}{8}
\end{array}\right]\otimes|2,1\rangle\langle 0,1|   \nonumber \\ 
+\left[\begin{array}{cc}
\frac{1}{8}& \frac{-i}{8}\\
\frac{-1}{8}    & \frac{i}{8}
\end{array}\right]\otimes |0,1\rangle\langle 2,1|+
\left[\begin{array}{cc}
\frac{1}{8}& \frac{-1}{8}\\
\frac{-1}{8}    & \frac{1}{8}
\end{array}\right]\otimes|0,1\rangle\langle 0,1|\, \nonumber \\
+\left[\begin{array}{cc}
\frac{1}{8}& \frac{i}{8}\\
\frac{-i}{8}    & \frac{1}{8}
\end{array}\right]\otimes |0,-1\rangle\langle 0,-1|+
\left[\begin{array}{cc}
\frac{-1}{8}& \frac{-i}{8}\\
\frac{-1}{8}    & \frac{-i}{8}
\end{array}\right]\otimes|-2,-1\rangle\langle 0,-1|   \nonumber \\ 
+\left[\begin{array}{cc}
\frac{-1}{8}& \frac{-1}{8}\\
\frac{i}{8}    & \frac{i}{8}
\end{array}\right]\otimes |0,-1\rangle\langle -2,-1|+
\left[\begin{array}{cc}
\frac{1}{8}& \frac{1}{8}\\
\frac{1}{8}    & \frac{1}{8}
\end{array}\right]\otimes|-2,-1\rangle\langle -2,-1|\, \nonumber\\
+\Phi(\left[\begin{array}{cc}
\frac{-1}{4}& \frac{-i}{4}\\
\frac{-1}{4}    & \frac{-i}{4}
\end{array}\right]\otimes|-1, 0\rangle\langle 1,0|+
\left[\begin{array}{cc}
\frac{-1}{4}& \frac{-1}{4}\\
\frac{i}{4}    & \frac{i}{4}
\end{array}\right]\otimes |1,0\rangle\langle -1,0|).\label{}
\end{eqnarray}

\vskip 0.2in
The probability distributions of the walk at the first three steps are computed below:
\begin{itemize}
  \item When $t=0$, $P(0)=1$;
  \item When $t=1$, $P(-1)=P(1)=\frac{1}{2}$;
  \item When $t=2$, $P(-2)=P(2)=\frac{1}{4}$, $P(0)=\frac{1}{2}$.
\end{itemize}

\section{Concluding remarks}

Discrete-time quantum walks have been formulated in terms of the state vectors in literature. An alternate formulation is introduced in this note. We describe quantum walks in the density operator picture. In the present framework, the action of the unitary channel on a density operator is described by the usual matrix multiplication. This approach offers, in the density operator picture, a mathematical model of unitary transformation of a quantum system in a directed graph.


\end{document}